\documentclass[aps,prl,twocolumn,showpacs]{revtex4}
\usepackage{graphicx}% Include figure files
\usepackage{dcolumn}% Align table columns on decimal point
\usepackage{bm}% bold math

\begin{document}

\title{Improved Limits on Long-Range Parity-Odd Interactions of the Neutron}
\author{E. G. Adelberger and T. A. Wagner}

\affiliation{Center for Experimental Nuclear Physics and Astrophysics, Box 354290,
University of
Washington, Seattle, WA 98195-4290}

\begin{abstract}
We show that a previous polarized $^3$He experiment at Princeton, plus E\"ot-Wash equivalence-principle tests, constrain exotic, long-ranged ($\lambda > 0.15\:$m) parity-violating interactions of neutrons at levels well below those inferred from a recent study of the parity-violating spin-precession of neutrons transmitted through liquid $^4$He. For $\lambda > 10^8\:$m the bounds on $g_Ag_V$ are improved by a 11 orders of magnitude.
 
\end{abstract}

\date{\today}
\maketitle
Yan and Snow\cite{ya:13} recently inferred bounds on the coupling strength, 
$g_A^n g_V^{\rm ^4He}$ 
%($g_V^{\rm ^4He}=2(g_V^{\rm p}+g_V^{\rm e}+g_V^{\rm n})$), 
of exotic, long-range, parity-violating interactions of neutrons from an experiment that studied the parity-violating spin-rotation of polarized neutrons transmitted through liquid $^4$He. %Here we show that s
Substantially tighter limits on several closely related quantities 
can be found by combining bounds on $|g_A^{\rm n}|^2$
and on $|g_V|^2$ set by previous experiments to obtain 
\begin{equation}
|g_A^n g_V|=\sqrt{|g_A^n|^2|g_V|^2}~.
\label{eq: sqrt}
\end{equation}
It is convenient to define
\begin{equation}
g_V^{\pm}=(g_V^{\rm p}+g_V^{\rm e}\pm g_V^{\rm n})/\sqrt{2}~,
\end{equation}
so that $g_V^{\rm ^4He}=2\sqrt{2}g_V^+$. 

We take our bounds on $|g_A^{\rm n}|$ from a Princeton optical-pumping experiment with polarized $^3$He detector and sources\cite{va:09,va:09a} 
that probed the neutron spin-spin interaction
\begin{equation}
V_{12}^{\sigma \cdot \sigma}=\frac{(g_A^n)^2}{4 \pi r}(\bm{\hat{\sigma}_1 \cdot \hat{\sigma}_2})e^{-r/\lambda}~,
\end{equation}
because the neutron in $^3$He carries most of the nuclear spin. 

\begin{figure}[t]
\hfil\scalebox{.45}{\includegraphics*[60pt,32pt][608pt,470pt]{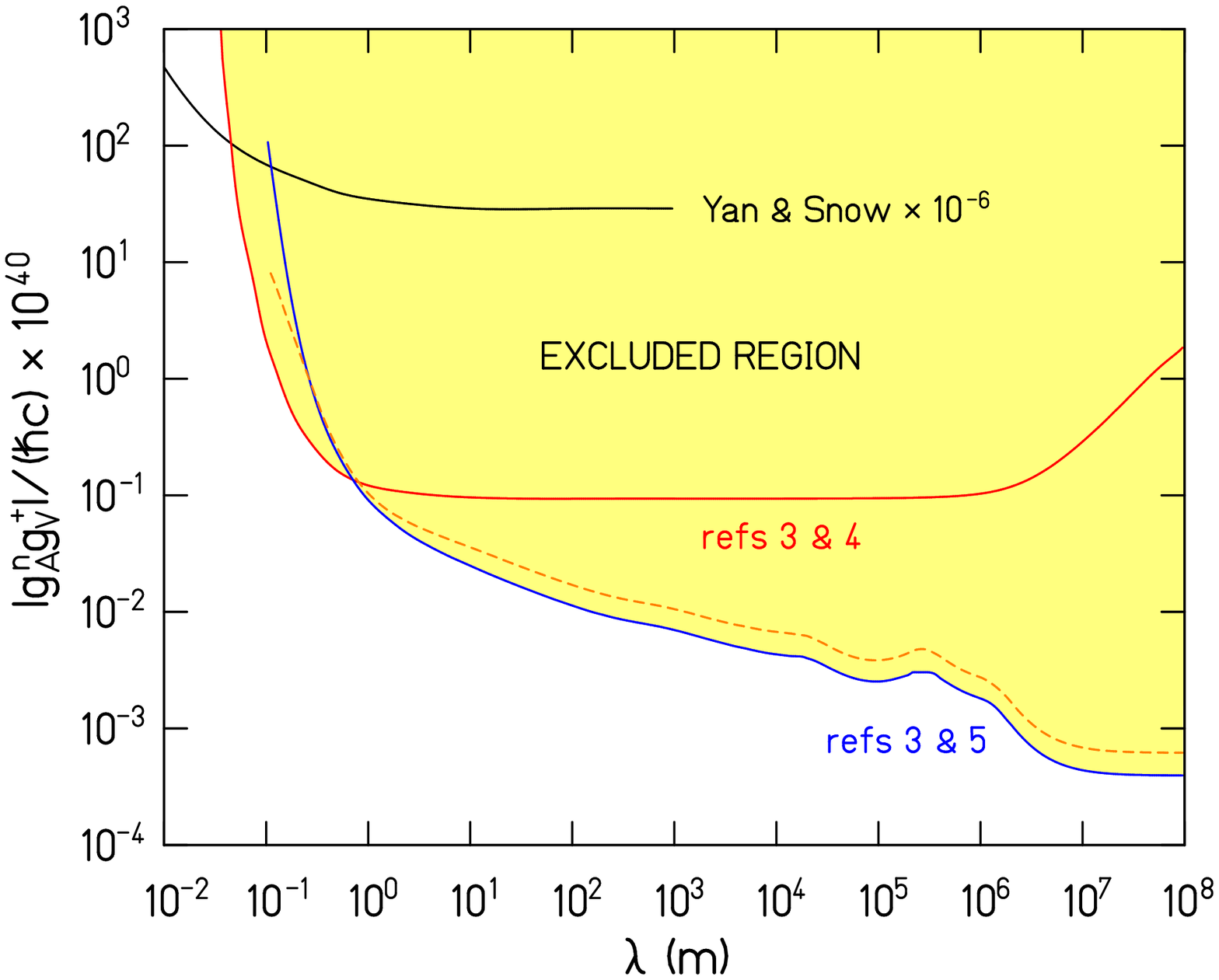}}\hfil
\caption{[Color online] Comparison of Yan and Snow's $1\sigma$ constraints on $|g_A^n g_V^+|$ \cite{ya:13} with those inferred from Princeton neutron spin-spin studies\cite{va:09} and E\"ot-Wash equivalence-principle tests with bodies falling toward a massive $^{238}$U laboratory source
\cite{sm:00} or in the field of the entire earth\cite{wa:12}. Our analysis of the E\"ot-Wash data assumes that $\tilde{q}^-=0$. Yan and Snow's upper bounds are divided by 6 orders of magnitude so that they can be displayed on the same scale. The dashed line shows our constraint
with no assumptions about the `charge' parameter $\tilde{\psi}$. 
\label{fig: gVgAplus}}
\end{figure}

\begin{figure}[t]
\hfil\scalebox{.45}{\includegraphics*[60pt,32pt][608pt,470pt]{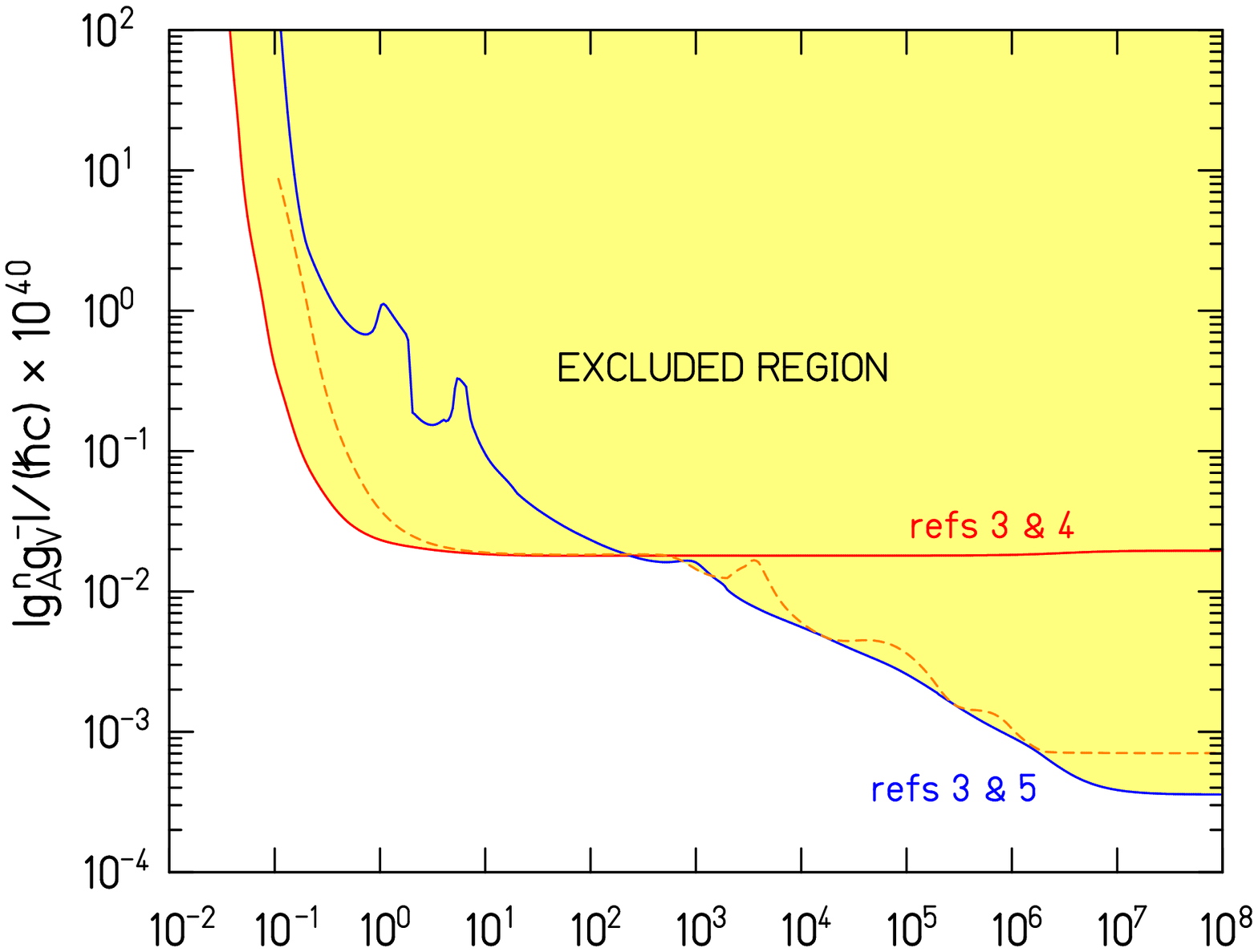}}\hfil
\caption{[Color online] $1\sigma$ constraints on $|g_A^n g_V^-|$ assuming that $g_V^+=0$. The  Ref.~\cite{wa:12} constraint is weaker and has more structure than in Fig.~\ref{fig: gVgAplus} because the earth consists largely of materials with $N\approx Z$. The undulations in the conservative bound (dashed line) occur where contributions to the source model (e.g.,  crust, mantle, or core)  with different compositions and densities change the value of $\tilde{\psi}^{\prime}$ that determines the greatest lower bound.
\label{fig: gVgAminus}}
\end{figure}

Our bounds on $|g_V|^2$ come from results of equivalence-principle tests\cite{sm:00,wa:12} %or short-distance inverse-square law tests\cite{ho:85,ka:07} that both 
that tightly constrain Yukawa interactions of the form
\begin{equation}
V_{12}=\frac{(g_V)^2}{4 \pi r}e^{-r/\lambda} 
=V_G(r)\: \tilde{\alpha}\left[\frac{\tilde{q}}{\mu}\right]_1 \left[\frac{\tilde{q}}{\mu}\right]_2 e^{-r/\lambda}~;
\label{eq: gVplusminus}
\end{equation}
where, in the second relation (conventionally used to analyze equivalence-principle results\cite{wa:12},
$V_G$ is the Newtonian potential, $\tilde{\alpha}$ is a dimensionless strength to be determined by experiment and a general vector `charge' of an atom with proton and neutron numbers $Z$ and $N$ can be parameterized as
\begin{equation}
\tilde{q}=\cos{\tilde{\psi}}[Z]+ \sin{\tilde{\psi}}[N]~,
\label{eq: q def}
\end{equation} 
where $\tilde{\psi}$ characterizes the vector charge with 
%is the number of baryons per amu and $\tilde{\alpha}$ is 
%are defined in Ref.~\cite{wa:12}. Note that 
\begin{equation}
\tan{\tilde{\psi}=\frac{g_V^n}{g_V^p+g_V^e}}~.
\label{eqn: psi def}
\end{equation}
Note that
$\tilde{q}^{\pm}$ correspond to `charge' parameters $\tilde{\psi}=\pm \pi/4$.  The results of this analysis are shown in Figs.~\ref{fig: gVgAplus} and \ref{fig: gVgAminus}. The $|g_V|^2$ constraint obtained from the Hoskins et al. inverse-square test\cite{ho:85} would be imperceptible in Figs. 1 and 2 because of the rapid weakening of the $|g_A^n|^2$ constraint\cite{va:09,va:09a} for $\lambda <0.2\:$m. 

%
%	Figure 1
%

\begin{figure} [h]
\hfil\scalebox{.45}{\includegraphics*[60pt,32pt][608pt,470pt]{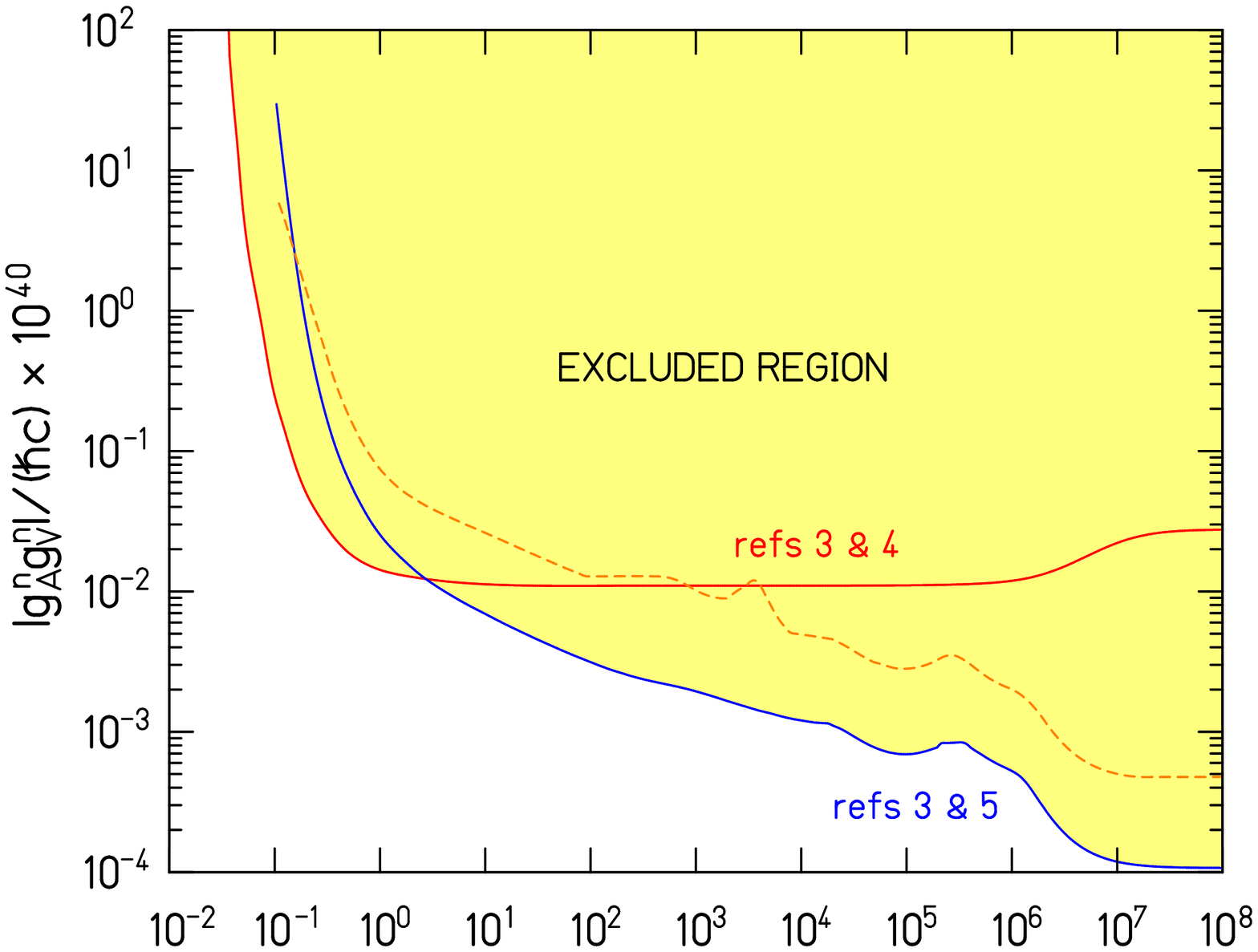}}\hfil
\caption{[Color online] Solid lines show $1\sigma$ constraints on $|g_A^n g_V^n|$ assuming that
$g_V^p+g_V^e=0$. The dashed line is a conservative constraint that makes no assumptions about $\tilde{\psi}$.
\label{fig: gVgAneutron}}
\end{figure}

The experimental results of Refs. \cite{va:09,va:09a,sm:00,wa:12} place especially tight bounds on $g_A^n g_V^n$, the strength  of a parity-violating neutron-neutron interaction. For this purpose we use Eqs. \ref{eq: gVplusminus} and~\ref{eq: q def} with  $\tilde{\psi}=\pi/2$ (i.e. $\tilde{q}=N$).
The differing sensitivities of the results in Figs.~\ref{fig: gVgAplus}, \ref{fig: gVgAminus} and \ref{fig: gVgAneutron} follow from the varying properties of the assumed charges. In Fig. 1, $\tilde{q}^+$ is proportional to the atomic mass number so that the $\tilde{q}/\mu$ ratio difference of the various equivalence-principle test-body pairs arises principally from the relatively small variation in $BE/Mc^2$ where $BE$ is the nuclear binding energy and $M$ the atomic mass. In Fig. 2 cancellation occurs between neutrons and protons because $N\approx Z$. The tightest limits occur in Fig. 3 because $\tilde{q}$ has no cancellations and $\tilde{q}/\mu \approx N/(Z+N)$ varies substantially for different test body materials.

We can do a completely general analysis by relaxing the assumptions made above about particular values of the `charge' parameter $\tilde{\psi}$. For example, to establish the most conservative bound on $g_V^n$ 
($\tilde{\psi}=\pi/2$) at a given value of $\lambda$ we fit the equivalence-principle constraints\cite{sm:00,wa:12} at that $\lambda$ for the entire range of $\tilde{\psi}^{\prime}$ values to obtain ${\tilde\alpha}(\lambda,\tilde{\psi}^{\prime})$, the functional dependence of $\tilde{\alpha}$ on $\tilde{\psi}$, and compute the conservative bound on $[g_V^n(\lambda)]^2$ from the greatest lower bound on
\begin{equation}
4 \pi G u^2 {\tilde\alpha}(\lambda,\tilde{\psi}^{\prime}) \cos^2({\tilde{\psi}-\tilde{\psi}^{\prime}})~,
\end{equation} 
where $G$ is the Newtonian constant and $u$ is the atomic mass unit.
This strategy requires equivalence-principle data with at least 2 different composition dipoles and 2 different attractors to avoid situations where either the charge of the attractor, or the charge-dipole of the pendulum, vanishes at a particular value of $\tilde{\psi}$.
The results are shown as dashed lines in Figs. 1-3.
%
%	Figure 2
%

The strategy employed above can also be used to find constraints on $|g_A^n g_V^{\pm}|$ for $\lambda < 1.5\times 10^{-2}$ m by taking $|g_V|^2$ from the inverse-square law tests of Hoskins et al.\cite{ho:85} and Kapner et al.\cite{ka:07} and $|g_A^n|^2$ from the cold-neutron experiment of Piegsa and Pignol\cite{pi:12}, but the  sensitivity of the cold-neutron work is not sufficient to give a result that is competitive with Yan and Snow's.  
%
%	Figure 3
%

We are indebted to Georg Raffelt for showing that tight bounds on exotic interactions can be obtained by combining the results of gravitational experiments and other data\cite{ra:12}.
This work was supported by NSF grant PHY969199.

\end{document}